\documentclass[twocolumn,aps,prd,superscriptaddress,nofootinbib,floatfix]{revtex4-2}

\usepackage{graphicx}
\usepackage{multirow}
\usepackage[colorlinks=true,citecolor=blue,urlcolor=blue,linkcolor=blue]{hyperref}

\usepackage{amssymb}
\usepackage{amsmath}

\usepackage{xcolor}
\definecolor{red}{rgb}{1, 0, 0}

\newcommand{\nn}{\nonumber}
\newcommand{\beq}{\begin{equation}}
\newcommand{\eeq}{\end{equation}}
\newcommand{\beqa}{\begin{eqnarray}}
\newcommand{\eeqa}{\end{eqnarray}}

\newcommand{\GeV}{{\rm GeV}}

\def\lqcd{\Lambda_{\rm QCD}}

\def\d{{\rm d}}

\newcommand{\Bbar}{\,\overline{\!B}{}}
\newcommand{\Dbar}{\,\overline{\!D}{}}
\newcommand{\Kbar}{\,\overline{\!K}{}}
\def\B0bar{\Bbar{}^0}
\def\D0bar{\Dbar{}^0}
\def\K0bar{\Kbar{}^0}

\newcommand{\hqs}{\ensuremath{\hat q^2}}
\newcommand{\hq}{\ensuremath{\hat q}}
\def\mth{\ensuremath{\hat m_\tau}}
\def\rt{\ensuremath{\rho_\tau}}

\def\OMIT#1{{}}

\makeatletter
\g@addto@macro\bfseries{\boldmath}
\makeatother

\begin{document}

\preprint{\vbox{\hbox{DESY 21-215}}}

\title{Theoretical predictions for inclusive \texorpdfstring{$B\to X_u \tau \bar\nu$}{} decay}

\author{Zoltan Ligeti}
\affiliation{Ernest Orlando Lawrence Berkeley National Laboratory,
University of California, Berkeley, CA 94720}

\author{Michael Luke}
\affiliation{Department of Physics, University of Toronto,
60 St.\ George Street, Toronto, Ontario, Canada M5S 1A7}

\author{Frank J.\ Tackmann}
\affiliation{Deutsches Elektronen-Synchrotron DESY, Notkestr.\ 85, 22607 Hamburg, Germany}

\begin{abstract}

With the expected large increase in data sets, previously not measured decays will be studied at Belle~II.  We derive standard model predictions for the $B\to X_u \tau\bar\nu$ decay rate and distributions.  
The region in the lepton energy spectrum where higher-dimension operators in the local OPE need to be resummed into the $b$-quark light-cone distribution function is a significantly greater fraction of the phase space than for massless leptons.  
The finite $\tau$ mass has the novel effect of shifting and squeezing how the distribution function enters the lepton energy spectrum.
We also derive new predictions for the $\tau$ polarization.

\end{abstract}

\maketitle

\newpage

\section{Introduction}

The more than $3\sigma$ deviation of the measured $B\to D^{(*)}\tau\bar\nu$ rates~\cite{Lees:2012xj, BaBar:2013mob, Aaij:2015yra, Huschle:2015rga, 
Belle:2017ilt, LHCb:2017rln, Belle:2019rba, Amhis:2019ckw, Bernlochner:2021vlv} from the standard model (SM) predictions motivates the study of all possible semileptonic decays with $\tau$
leptons in the final state, both experimentally and theoretically.  Comparisons
of measured spectra and rates to different hadronic final states can give
information on the structure of contributing four-fermion operators. 
Comparisons of $b\to c\ell\bar\nu$ and $b\to u\ell\bar\nu$ decays give
constraints on the flavor structure of beyond standard model scenarios at play. 

In this paper we study the inclusive decay $B\to X_u\tau\bar\nu$, which has been much less explored theoretically.  Precise predictions for
this decay are naturally interesting as a signal channel to measure in the future.  In the near term, reliably modelling this decay as a background is interesting both to SM measurements and analyses aimed at more precisely measuring $R(D^{(*)})$ and clarifying the current tension with the SM.  The Belle Collaboration set the first bound on a $b\to u\tau\bar\nu$
mediated decay, ${\cal B}(B\to\pi\tau\bar\nu) < 2.5\times
10^{-4}$~\cite{Hamer:2015jsa}, at a level several times higher than SM predictions, and recent theoretical studies~\cite{Bernlochner:2015mya, Bhatta:2020yvb, Bernlochner:2021rel} also focused on exclusive decays.

Inclusive semileptonic decays of hadrons containing a heavy quark allow for a
systematic expansion of nonperturbative effects in powers of
$\lqcd/m_Q$~\cite{Chay:1990da}.  The inclusive decay rates computed in the
$m_Q\gg\lqcd$ limit coincide with the free-quark decay rate, while
corrections of order $\lqcd/m_Q$ vanish~\cite{Chay:1990da, Bigi:1992su}.  The
leading nonperturbative corrections are of order $\lqcd^2/m_Q^2$ and depend on only
two hadronic quantities, $\lambda_1$ and $\lambda_2$, which describe certain forward matrix elements of
local dimension-five operators.  These corrections have been computed for a
number of processes~\cite{Bigi:1993fe, Blok:1993va, Manohar:1993qn, Falk:1993dh, Koyrakh:1993pq, Balk:1993sz, Falk:1994gw}. For $B\to X_u\tau\bar \nu$ decay, expressions for the total rate and leptonic $q^2$ spectra are straightforward to derive by taking the $m_q\to 0$ limit of the $B\to X_c\tau\bar\nu$ results~\cite{Falk:1994gw,Ligeti:2014kia}, but this limit is singular for the lepton energy spectrum and has not been given in the literature. Similarly, the perturbative ${\cal O}(\alpha_s)$ corrections to the total $B\to X_u\tau\bar\nu$ semileptonic decay rate~\cite{Hokim:1983yt}, the dilepton $q^2$ spectrum~\cite{Czarnecki:1994bn}, and the doubly differential $\d\Gamma/\d q^2 \d y$ spectrum~\cite{Jezabek:1996db,Jezabek:1997rk} are known analytically.
However, no closed form expressions have thus far been derived for the ${\cal O}(\alpha_s)$ corrections to the $\tau$ lepton energy spectrum.  We present the results of the local OPE to ${\cal O}(\lqcd^2/m_b^2,\, \alpha_s)$ in Sec.~\ref{sec:local}.

Phase space regions in inclusive $B\to X_u e\bar\nu$ decay, when kinematic cuts restrict the invariant mass of the hadronic final state to be small (i.e., $m_X < m_D$), are relevant
for the determination of $|V_{ub}|$.  Decay rates in such regions are subject to large corrections, both
perturbative and nonperturbative.  In the region near maximal lepton energy the OPE
breaks down and a resummation of the series of leading nonperturbative
corrections is required~\cite{Neubert:1993ch, Bigi:1993ex}.  The lepton energy spectrum in a region of width $\Delta E_\ell \sim \lqcd$ near the endpoint is determined by a nonperturbative $b$-quark distribution function in the $B$ meson. Similarly, the local OPE for $B\to X_u\tau\bar\nu$ breaks down near the endpoint of the $\tau$ energy spectrum; however, since in $B\to X_u \tau \bar\nu$ decay, $m_\tau < E_\tau < (m_B^2 + m_\tau^2)/(2m_B)$ amounts to $1.78\,\GeV < E_\tau < 2.94\,\GeV$, the distribution function is important over a much greater fraction of the available phase space than in $B\to X_u e \bar\nu$, where $0 < E_e < m_B/2$. We consider the effects of the $b$-quark distribution function in Sec.~\ref{sec:shapefn} and explore its effect on the spectrum.
Since the distribution of the measurable $\tau$ decay products (e.g., the
charged lepton energy) are sensitive to the $\tau$ polarization, we also present
results for decays to each polarization state.

To appreciate the mass suppressions in the decay rates, simply using the ${\cal
O}(\lqcd^2/m_b^2)$~\cite{Koyrakh:1993pq, Balk:1993sz, Falk:1994gw} and ${\cal O}(\alpha_s)$ contributions~\cite{Hokim:1983yt, Falk:1994gw} in the $1S$
scheme~\cite{Hoang:1998ng, Hoang:1998hm, Hoang:1999zc}, one
finds~\cite{Hoang:1998hm}
\beq\label{ltauratio}
\frac{\Gamma(B\to X_u\ell\bar\nu)}{\Gamma(B\to X_u\tau\bar\nu)}\,
  = 2.97 \,, \quad
\frac{\Gamma(B\to X_c\ell\bar\nu)}{\Gamma(B\to X_c\tau\bar\nu)}\,
  = 4.50 \,,
\eeq
Thus, the suppression of the rate due to finite $m_\tau$ is less strong in $b\to u$ than in $b\to c$ decays.
Correspondingly, the suppression due to finite $m_c$ is clearly greater in $B\to \tau$ than in $B\to e$ semileptonic decays,
\beqa\label{ucratio}
\frac{\Gamma(B\to X_u\tau\bar\nu)}{\Gamma(B\to X_c\tau\bar\nu)}\,
  \frac{|V_{cb}|^2}{|V_{ub}|^2} &=& 3.13\,, \nn\\*
  \frac{\Gamma(B\to X_u\ell\bar\nu)}{\Gamma(B\to X_c\ell\bar\nu)}\,
  \frac{|V_{cb}|^2}{|V_{ub}|^2} &=& 1.83 \,.
\eeqa

\section{local ope results}
\label{sec:local}

\subsection{Nonperturbative Corrections}

The inclusive $B\to X_q\,\ell\bar\nu$ decay ($q=u,c$\,; $\ell=e,\mu,\tau$) has been
considered to order $1/m_b^2$ in the heavy quark expansion~\cite{Koyrakh:1993pq, Balk:1993sz, Falk:1994gw}, including effects of the finite lepton mass.  For
$m_q=0$ the lepton energy spectrum becomes singular, and the limit must be taken with care. We find for $B\to X_u \tau\bar\nu$ decay,%
\footnote{The results in this section apply, with obvious changes of hadron masses and matrix elements, to inclusive $B_c \to X_c\tau\bar\nu$ decay, just like exclusive $B_c$ decays can be calculated using HQET methods~\cite{Jenkins:1992nb}.  Treating charm as a heavy quark, the $B_c$ has a size parametrically smaller than $\lqcd$, and the $b$ quark distribution function in $B_c$ is in principle calculable in NRQCD.  This decay might be observable in the tera-$Z$ phase of a future $e^+e^-$ collider.}
\beqa\label{expl}
\frac1{\Gamma_u} \frac{\d\Gamma}{\d y} &=& 
  2\sqrt{y^2-4\rt}\, \bigg[3y-2y^2-4\rt+3\rt y
  + \frac{\lambda_2}{m_b^2}\,6y \nn\\*
&&{} + \frac{\lambda_1 + 3\lambda_2}{3m_b^2}\, (5y^2-14\rt)
  \bigg]\, \theta(1+\rt-y) \nn\\
&-& \bigg[ \frac{\lambda_1}{3m_b^2}\, (1+\rt)
  + \frac{\lambda_2}{m_b^2}\, (11-5\rt) \bigg] \nn\\
&&{} \times (1-\rt)^3\, \delta(1+\rt-y) \nn\\
&-& \frac{\lambda_1}{3m_b^2}\, (1-\rt)^5\, \delta'(1+\rt-y) \,,
\eeqa
where we use the dimensionless variables
\beq
y = \frac{2E_\tau}{m_b}\,, \qquad \hqs = \frac{q^2}{m_b^2}\,, \qquad
  \rt = \frac{m_\tau^2}{m_b^2}\,,
\eeq
and
\beq
\Gamma_u = \frac{|V_{ub}|^2\,G_F^2\,m_b^5}{192\,\pi^3}\,.
\eeq
This agrees with the more complicated expression given in Ref.~\cite{Balk:1993sz}.
Here $\lambda_1$ and $\lambda_2$ are matrix elements in the heavy quark 
effective theory (HQET), defined by
\begin{eqnarray}\label{ldef}
\frac1{2m_B}\langle B|\, \bar b_v\,(iD)^2\, b_v |B\rangle 
  &=& 2\,\lambda_1 \,, \nonumber\\
\frac1{2m_B}\langle B|\, \frac{g_s}2\,\bar b_v\,\sigma_{\mu\nu} G^{\mu\nu}\, b_v 
  |B\rangle &=& 6\,\lambda_2 \,,
\end{eqnarray}
and $b_v$ is the heavy $b$-quark field of HQET~\cite{Georgi:1990um} with velocity $v$.

The $\tau$ can have spin up ($s=+$) or spin down ($s=-$)
relative to the direction of its three-momentum, and it is convenient to
decompose the corresponding decay rates as
\beq
\Gamma(B \to X_u\, \tau(s=\pm)\, \bar\nu) = \frac12\, \Gamma \pm \tilde\Gamma \,.
\eeq
The rate, summed over the tau polarizations, is given by $\Gamma$, while
the average tau polarization is $A_{\rm pol} = 2\tilde\Gamma / \Gamma$.  
The $\tau$ polarization gives complementary sensitivity to BSM physics~\cite{Kalinowski:1990ba}. We obtain for its lepton energy dependence,
\beqa\label{ypol}
\frac1{\Gamma_u} \frac{\d\tilde\Gamma}{\d y} &=& - (y^2-4\rt) 
  \bigg[ 3-2y+\rt \nn\\*
&&{}\quad +\frac{6\lambda_2}{m_b^2}
  + \frac{\lambda_1+3\lambda_2}{3m_b^2} 5y \bigg]\, \theta(1+\rt-y) \nn\\
&+& \bigg[ \frac{\lambda_1}{6m_b^2}\, (1-3\rt)
  + \frac{\lambda_2}{2m_b^2}\, (11-5\rt) \bigg] \nn\\
&&{}\quad \times (1-\rt)^3\, \delta(1+\rt-y) \nn\\
&+& \frac{\lambda_1}{6m_b^2}\, (1-\rt)^5\,  \delta'(1+\rt-y) \,.
\eeqa
Note that for
$\rho_\tau=0$, $-2\, \d\tilde\Gamma = \d\Gamma$, since the massless lepton is purely left-handed.
Angular momentum conservation in $B\to X_u\tau\bar\nu$ implies that the $\tau$ polarization is fully left-handed at maximal $E_\tau$.
This holds at the parton level to all orders in $\alpha_s$, and our results indeed satisfy it at order $\alpha_s^0$ and order $\alpha_s^1$; i.e., $\Gamma/2 = - \tilde\Gamma$ at $y=1+\rt$.
However, the power-suppressed terms that start at order $\lqcd^2/m_b^2$ incorporate nonperturbative corrections between the $E_\tau$ endpoint at the parton level and at the hadron level.  As a result, the physical rate at maximal $E_\tau$ vanishes (it is nonzero at the parton level).  In a small region very close to the endpoint the most singular terms of the form $\lambda_1\, \delta'(1+\rt-y)$ are the most important, and these also obey the $\Gamma/2 = - \tilde\Gamma$ relation.

For $\d\Gamma/\d \hqs$, the $m_q\to 0$ limit the of $B\to X_c\tau\bar\nu$ expression is smooth, which gives the known result~\cite{Ligeti:2014kia},
\beqa\label{q2specU}
\frac1{\Gamma_u} \frac{\d\Gamma}{\d\hqs} &=& \frac{(\hqs-\rt)^2}{\hq^6}
  \bigg\{ \bigg( 1+\frac{\lambda_1}{2m_b^2}\bigg) 2 (1-\hqs)^2 \nn\\
&&{} \times \Big[ \hqs(1 + 2\hqs) + \rt(2+\hqs) \Big] \\
&+& \frac{3\lambda_2}{m_b^2}
\Big[\hqs(1-15\hq^4+10\hq^6) + \rt(2-3\hqs+5\hq^6) \Big]\! \bigg\}.\nn
\eeqa
Integrating over phase space, the $B\to X_u\tau\bar\nu$ rate is
\beqa\label{rate}
\frac{\Gamma}{\Gamma_u}&=& \bigg( 1+\frac{\lambda_1}{2m_b^2}\bigg)
  \big( 1-8\rt+8\rt^3-\rt^4-12\rt^2\ln\rt \big) \\
&& - \frac{3\lambda_2}{2m_b^2}
  \big( 3-8\rt+24\rt^2-24\rt^3+5\rt^4+12\rt^2\ln\rt \big)\,, \nn
\eeqa
and the polarization is given by
\beqa\label{pol}
\frac{\tilde \Gamma}{\Gamma_u} &=& - \frac{(1-\mth)^3}2 \bigg[ \frac{(1-\mth)^2}3\,
  \big(3+15\mth+5\mth^2+\mth^3 \big) \nn\\
&&{} + \frac{\lambda_1}{6m_b^2}\,
  (1+\mth)^3\, (3+\mth^2) \\
&& - \frac{\lambda_2}{2m_b^2}
  \big(9 +27\mth +70\mth^2 +10\mth^3 -15\mth^4-5\mth^5 \big) \bigg] , \nn
\eeqa
where $\mth = \sqrt{\rt}$.

\subsection{Perturbative Corrections}

Analytic results for the doubly differential $\d\Gamma/\d q^2 \d y$ spectra (including the $\tau$ polarization dependence) were given in Refs.~\cite{Jezabek:1996db,Jezabek:1997rk}\,\footnote{We corrected some typos in the $m_c\to 0$ limit in these references.}, but only numerical results were presented for the $\tau$ energy spectrum. Integrating the doubly differential spectra over $q^2$ gives the charged lepton energy spectra for both unpolarized and polarized $\tau$ leptons. In the unpolarized case, writing
\newpage
\begin{equation}\label{partonspect}
\frac{1}{\Gamma_0}\frac{\d\Gamma_\tau}{\d y} = \Big[F_0(y)-\frac{\alpha_s C_F}{2\pi} F_1(y) \Big]\, \theta(1+\rt-y) \,,
\end{equation}
where $C_F = 4/3$, we find
\begin{equation}
F_0(y) = 2\sqrt{y^2-4 \rt}\, \big[(3-2y)y+\rt(3y-4)\big] \,,
\end{equation}
and
\begin{widetext}
\begin{equation}\label{pert}
\begin{aligned}
F_1(y) ={}& F_0(y) \left[\text{Li}_2(\tau_+) + \text{Li}_2(\tau_-) + 4 Y_p^2\right]
  +\left(6 y^2-4 y^3+6 \rt y^2-12 \rt ^2\right) \left[\text{Li}_2(\tau_+)-\text{Li}_2(\tau_-)\right] \\
& -2Y_p\left(\frac{5 \rt ^3}{3} + \rt(7 y^2-6 y+7) - 6 y^3+10 y^2 + \rt ^2(4 y-23) + 6y-\frac{41}{3}\right)   \\
& + \sqrt{y^2-4 \rt} \left(-\frac{34 y^2}{3} 
  + \rt \left(15 y-\frac{74}{3}\right)+24 y-6\right) \ln(1-y+\rt) \\
&+ \sqrt{y^2-4 \rt} \left(21 \rt ^2+\frac{1}{6} \left[\left(86-16 \pi ^2\right) y^2 +\left(24 \pi ^2-153\right) y+82\right]+\frac{\rt}{6} \left(24 \pi ^2 y -167 y-32 \pi ^2+64\right) \right) ,
\end{aligned}
\end{equation}
\end{widetext}
where $Y_p= \frac12 \ln[(1-\tau_+)/(1-\tau_-)]$ is the rapidity of all decay products (combined) against which the $\tau$ recoils,~and
\begin{equation}\label{tpm}
\tau_\pm = \frac12 \left(y\pm\sqrt{y^2-4\rt}\right) .
\end{equation}
Similarly, defining the polarization dependence of the lepton energy spectrum as
\begin{equation}
\frac{\d\Gamma^\pm_\tau}{\d y} = \frac12\frac{\d\Gamma_\tau}{\d y}\pm \frac{\d\tilde\Gamma_\tau}{\d y} \,,
\end{equation}
we write the polarization dependence of the rate to produce a $\tau$ lepton as
\begin{equation}\label{partonpol}
\frac{1}{\Gamma_0}\frac{\d\tilde\Gamma_\tau}{\d y} = \Big[ \tilde F_0(y) - \frac{\alpha_s C_F}{2\pi} \tilde F_1(y) \Big]\, \theta(1+\rt-y) \,.
\end{equation}
At tree level,
\begin{equation}
\tilde F_0(y) = \left(y^2-4\rt\right)(2y-3-\rt) \,,
\end{equation}
while at one loop,
\begin{widetext}
\begin{align}
\tilde F_1(y) ={}& \tilde F_0(y)\left[\text{Li}_2(\tau_+) + \text{Li}_2(\tau_-) + 4  Y_p^2\right]
  +\frac{12 \rt ^2-\rt \left(y^3+6 y^2-6 y\right)+2 y^4-3 y^3}{\sqrt{y^2-4 \rt }}
  \left[\text{Li}_2(\tau_+)-\text{Li}_2(\tau_-)\right] \nn\\
& - \frac{Y_p}{3\sqrt{y^2-4 \rt}} \bigg[ \rt \left(y^3-210 y^2+405 y-260\right) 
  + \left(18 y^4-12 y^3-36 y^2+41 y\right)
  + \rt^3 (y-24) + \rt^2 (93 y-12) \bigg]  \nn\\
& +\left[\frac{34 \rt ^2}{3}+\rt  \left(-\frac{23 y^2}{6}-\frac{53 y}{3}+30\right)+\frac{17 y^3}{3}-9 y^2\right] \ln(1-y+\rt)
+\frac{\rt ^3}{3} + \rt ^2 \left(-\frac{11 y}{6}+\frac{8 \pi ^2}{3}-21\right) \nn\\
&+ \frac{\rt}{12} \left(-8 \pi ^2 y^2+149 y^2-64 \pi ^2 y-48 y+96 \pi ^2+32\right) 
  + \frac{1}{12} \left(16 \pi ^2 y^3-86 y^3-24 \pi ^2 y^2+153 y^2-82 y\right)
.
\end{align}
\end{widetext}

\section{The lepton energy endpoint region}
\label{sec:shapefn}

Near the endpoint of the lepton energy spectrum $y\sim 1+\rho_\tau$, a class of higher-order terms in the local OPE in Eq.~(\ref{expl}) is no longer suppressed, and instead the differential rate is given by a nonlocal OPE in terms of the light-cone momentum distribution function of the $b$ quark~\cite{Neubert:1993um,Neubert:1993ch,Falk:1993vb,Mannel:1994pm,Bigi:1993ex,Bigi:1994it}.

This endpoint region has been extensively studied in the context of massless leptons. It is straightforward to extend this to nonzero $\tau$ mass. At the parton level the lepton energy endpoint is determined by the $\theta$ function
\begin{equation}
    \theta\big((p_b-p_\tau)^2\big)=\theta\left( m_b^2+m_\tau^2- 2\,p_\tau\cdot p_b\right).
\end{equation}
Writing
\begin{equation}
    p_\tau^\mu=\frac{m_b}2 \big(\tau_-\, n^\mu + \tau_+\, \bar n^\mu\big)\,,
\end{equation}
where $\tau_\pm$ are given in Eq.~(\ref{tpm}), defines the light-like vectors $n^\mu$ and $\bar n^\mu=2v^\mu-n^\mu$. Taking
$p_b^\mu=m_b v^\mu+k^\mu$,
expanding in powers of $k^\mu/m_b$ and applying the HQET onshell condition $k\cdot v=0$, the $\theta$ function becomes
\begin{equation}\label{thetafn}
    \theta\left(1+\rho_\tau-y + \frac{k\cdot n}{m_b}\, \sqrt{y^2-4 \rho_\tau}+{\cal O}(k^2)\right).
\end{equation}
Over most of the spectrum, the ${\cal O}(k\cdot n)$ term may be neglected at leading order in $1/m_b$ and we recover the OPE result in Eq.~(\ref{expl}).  However, when $E_\tau$ is near the partonic endpoint, i.e., $1+\rt-y = {\cal O}(\lqcd/m_b)$, $p_b-p_\tau$ approaches a light-like vector in the $n$ direction. In this region the ${\cal O}(k\cdot n)$ term is the same order as the leading term, and so must be included in the leading-order expression.
Defining
\begin{equation}\label{deltadef}
    \Delta\equiv \frac{1+\rho_\tau-y}{1-\rho_\tau}\,,
\end{equation}
taking $\Delta\sim {\cal O}(\lqcd/m_b)$, and expanding \eqref{thetafn} in powers of $\Delta$ then gives 
\begin{equation}
\theta\bigg(\Delta+\frac{k\cdot n}{m_b}\bigg) + {\cal O}(\lqcd/m_b)\,.
\end{equation}
Comparing with the $\rho_\tau\to0$ limit, the nonzero $\tau$ mass shifts the endpoint of the lepton spectrum and squeezes it by a factor of $1-\rho_\tau$.
This is also reflected by the fact that the lepton energy endpoint changes between the parton- and hadron-level kinematics, at leading order, by $(1-\rt)\,\bar\Lambda/2$,
where $m_B = m_b + \bar\Lambda + {\cal O}(\lqcd^2/m_b)$.

At the hadron level, matrix elements of the $\theta$ function may be expressed as an integral over the light-cone momentum distribution function of the $b$ quark in the $B$ meson,
\begin{equation}
    f(\omega,\mu)= \frac1{2m_B}\, \langle B| \bar b_v\delta(\omega+i D\cdot n)b_v|B\rangle \,.
\end{equation}
Following \cite{Ligeti:2008ac}, it is convenient to define the nonperturbative function $F(k)$ via the convolution
\begin{equation}\label{fconv}
    f(\omega,\mu)=\int \d k\, C_0(\omega-k,\mu)\, F(k)\,,
\end{equation}
where, at one loop~\cite{Bauer:2003pi},
\begin{equation}
    C_0(\omega,\mu) = \delta(\omega) - \frac{\alpha_s C_F}{4\pi} \Bigg( \frac{\pi^2}{6} \delta(\omega) + \frac4\mu \bigg[\frac\mu\omega\bigg]_+ + \frac8\mu \bigg[\frac{\ln\frac\omega\mu}{\omega/\mu}\bigg]_+ \Bigg).
\end{equation}
The convolution \eqref{fconv} factors out the perturbative corrections to the parton-level matrix element of $f(\omega)$. With this definition, $F(k)$ is a nonperturbative function with support from $k=-\bar\Lambda$ to $k=\infty$, whose moments are related to the matrix elements of local operators. The $\tau$ energy spectrum may then be written in the endpoint region as the convolution
\begin{equation}\label{shapeconv}
\frac1{\Gamma_\tau} \frac{\d\Gamma_\tau}{\d y} = \int \d\omega\, G_\tau\bigg(\Delta-\frac\omega{m_b}\bigg)  F(\omega) 
  + {\cal O}(\Delta,\lqcd/m_b) \,,
\end{equation}
where $G_\tau(x)$ is obtained by expanding the parton level perturbative results \eqref{pert} in the limit $\Delta\to 0$,
\begin{align}\label{Gtau}
G_\tau(x) = \theta(x)\, & \bigg\{ 1 - \frac{\alpha_s C_F}{2\pi} \bigg[\ln^2 x \\ 
& + \bigg(\frac{31}6 - 2 \ln(1-\rho_\tau)\bigg)\ln x 
  + C(\rho_\tau) \bigg]\bigg\} \,, \nn
\end{align}
and $C(\rt) = \pi^2+5/4 + \rt (\pi^2-6) + {\cal O}(\rt^2)$. Note that in Eq.~(\ref{Gtau}) the $m_\tau$ dependent terms at ${\cal O}(\alpha_s)$ are small corrections: $C(\rt)/C(0)$ is within 5\% of unity, and the $2\ln(1-\rt)$ term is less than a 6\% correction relative to the ``31/6" term. $G_\tau(x)$ therefore has very weak $\rho_\tau$ dependence: none at tree level, and only about $5\%\times\alpha_s C_F/(2\pi)$ at one loop. The large difference between the shapes arises almost entirely from the kinematic rescaling in Eq.~(\ref{deltadef}).  SCET techniques may be used to sum logarithms of $\Delta$ in this expression (as in Refs.~\cite{Bauer:2003pi} and \cite{Ligeti:2008ac}), but this is beyond the scope of this paper or the accuracy we desire.

\begin{figure*}[tbh!]
\includegraphics[width=\textwidth]{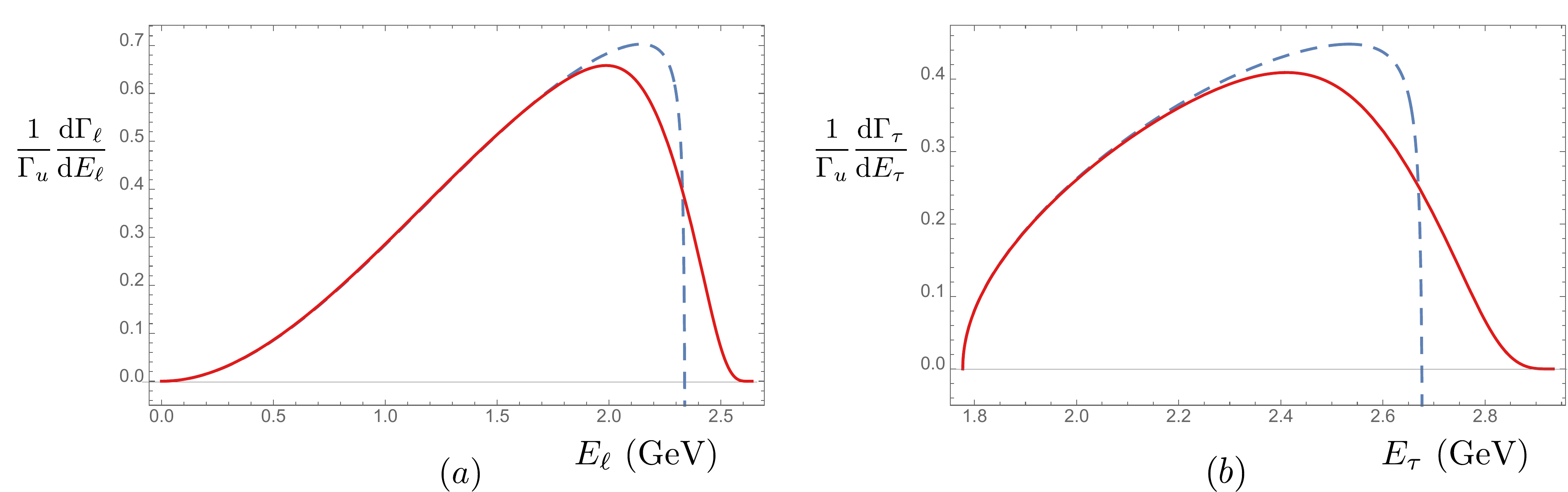}
\caption{The $B\to X_u\ell\bar\nu$ lepton energy spectrum for (a) $\ell=e,\, \mu$ and (b) $\ell=\tau$ in the parton model (blue, dashed), and incorporating the leading order $b$-quark distribution function (red, solid).}
\label{fig:tauspec}
\end{figure*}

\begin{figure}[tbh]
\includegraphics[width=\columnwidth]{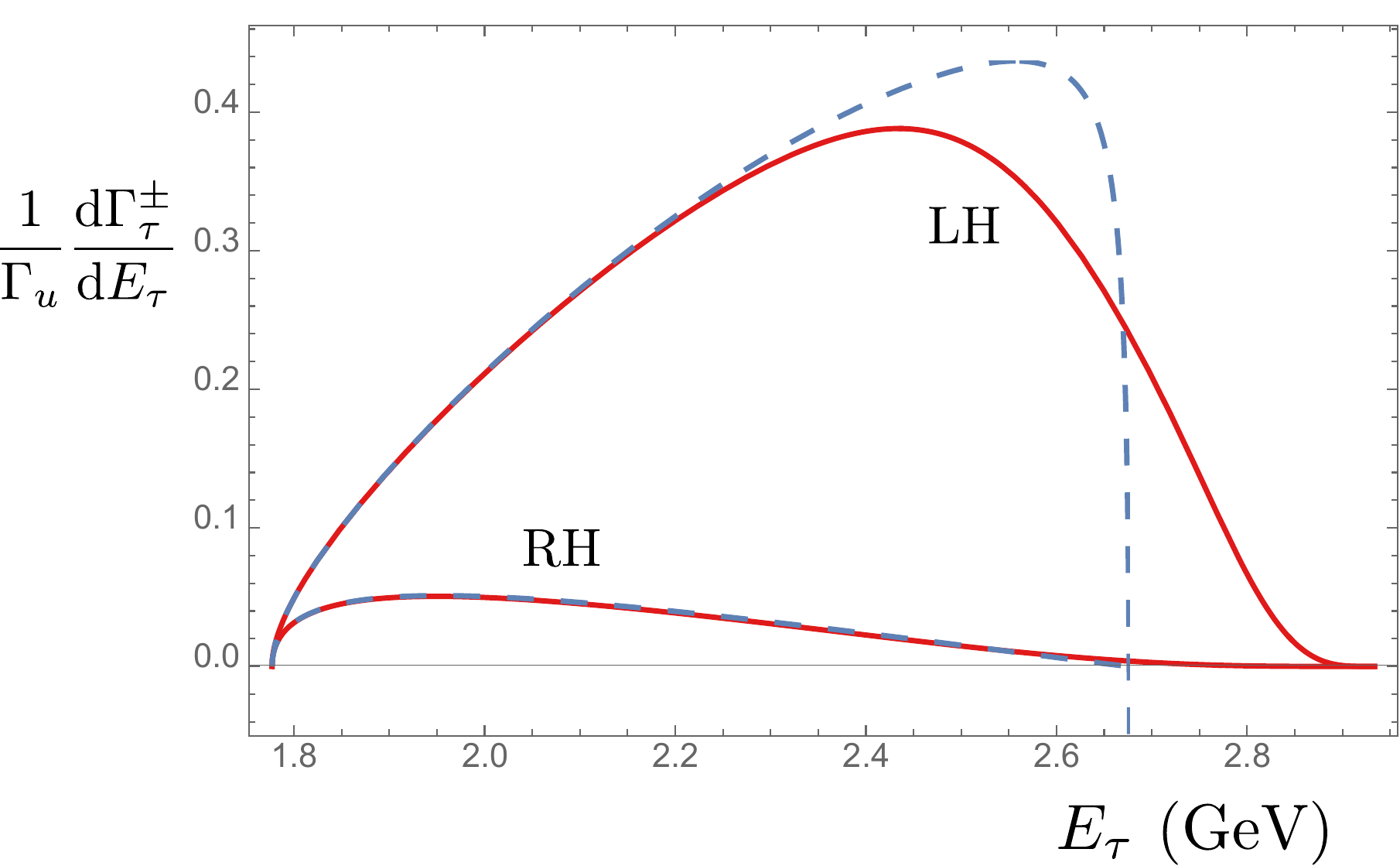}
\caption{The $\tau$ energy dependence of its polarization in $B\to X_u\tau\bar\nu$ in the parton model (blue, dashed), and incorporating the leading order $b$-quark distribution function (red, solid).}
\label{fig:tauspec2}
\end{figure}

The expression (\ref{shapeconv}) is only valid in the region $\Delta\sim\lqcd/m_b$; in order to have an expression which smoothly interpolates with the local OPE away from the endpoint, it is convenient instead to incorporate distribution function effects by redefining the $b$-quark mass $m_b\to m_b^\prime=m_b+k\cdot n$ \cite{Neubert:1993um,Neubert:1993ch}.  Writing $p_b^\mu=m_b^\prime v^\mu+ k^{\prime\mu}$, where $k^{\prime\mu}=k^\mu-k\cdot n\, v^\mu$, the residual momentum $k^{\prime\mu}$ satisfies $k^\prime\cdot n=0$, and so the effects of nonzero $k\cdot n$ are automatically incorporated into the leading-order spectrum with this mass definition. The $\tau$ energy spectrum in the endpoint region may then be written as the convolution
\begin{equation}\label{shapefn}
    \frac{\d\Gamma_\tau}{\d E_\tau} = 2\int \d\omega\, 
    \frac1{m_b}\frac{\d\Gamma_\tau}{\d  y}(y^\prime, \rho_\tau^\prime)\, F(\omega)\,,
\end{equation}
where we have defined the scaled variables
\begin{equation}
    y^\prime \equiv \frac{2E_\tau}{m_b-\omega}\,, \qquad 
    \rho_\tau^\prime\equiv\frac{m_\tau^2}{(m_b-\omega)^2}\,,
\end{equation}
and $\d\Gamma/\d y$ is the parton level spectrum in Eq.~(\ref{partonspect}). An analogous formula holds for the polarized spectrum Eq.~(\ref{partonpol}).  For simplicity, we have written the prefactor in Eq.~(\ref{shapefn}) as $1/m_b$, not $1/(m_b-\omega)$, since the difference is higher order everywhere in the spectrum.
In this form, Eq.~(\ref{shapefn}) includes subleading terms suppressed by powers of $\Delta$ in the endpoint region, but which are leading order when $\Delta$ is not small, so are required to reproduce the local OPE away from the endpoint. 

$F(k)$ has been extracted from the measured $B\to X_s\gamma$
spectra by the SIMBA collaboration~\cite{Bernlochner:2020jlt}. 
At leading order in $\lqcd/m_b$, it can be used to make predictions for $B\to X_u\ell\bar\nu$ decays.
Figure~\ref{fig:tauspec} shows the $B\to X_u\ell\bar\nu$ lepton spectra for $\ell=e$ and $\ell=\tau$ in the parton model and including the effects of the $b$-quark distribution function.  It is clear from this plot that the distribution function is indeed important in a greater fraction of the $\tau$ energy spectrum than in the massless lepton channels; the fraction of the lepton energy spectrum where the distribution function is important is enhanced by $(1-\rt)/(1-\sqrt\rt)^2 \sim 2.2$.
Figure~\ref{fig:tauspec2} shows the $E_\tau$ spectra separately for left- and right-handed $\tau$ leptons in $B\to X_u\tau\bar\nu$.  The average $\tau$ polarization, including order $\alpha_s$ and $\lqcd^2/m_b^2$ corrections, is $2\tilde\Gamma/\Gamma = -0.77$.

\section{Conclusions}

We presented theoretical predictions for inclusive $B\to X_u\tau\bar\nu$ decay. We derived previously unknown results at order $\lqcd^2/m_b^2$ and analytic expressions for the order $\alpha_s$ corrections for the $\tau$ energy spectrum and polarization. We also incorporated the effects of the $b$-quark light-cone distribution function to the case of nonzero lepton mass. Due to the suppressed kinematic range, the $b$-quark distribution function is more important in determining the lepton energy spectrum in $B\to X_u\tau\bar\nu$ than in $B\to X_u e\bar\nu$ decay.

It will probably take many ab$^{-1}$ of data at Belle~II to have sensitivity to 
$B\to X_u \tau\bar\nu$.  While it is clearly a challenging decay to measure, the rate according to
Eqs.~(\ref{ltauratio}) and (\ref{ucratio}) is only about 3 times smaller than
$B\to X_u e\bar\nu$, and about $|V_{cb}|^2/(3\, |V_{ub}|^2)$ times smaller than $B\to
X_c \tau\bar\nu$.  One may, for example, try to utilize the fact that electrons
or muons from the $\tau$ decay with maximal allowed energies correspond to the
most energetic $\tau$ leptons.  We hope that Belle~II will be able to make
measurements of this decay.

\acknowledgments

This paper is dedicated to the memory of Sheldon Stone, with whom we had countless inspiring and entertaining discussions, e.g., related to the papers~\cite{CLEO:2000tsw, Ligeti:2001dk, CLEO:2001vfw}; he'd surely be appalled knowing how long this one took to get out. 
We thank Mark Wise for helpful conversations about these decays in the previous millennium, and Florian Bernlochner and Aneesh Manohar in this one.
ZL thanks the Aspen Center for Physics (supported by the NSF Grant PHY-1607611) for hospitality while some of this work was carried out.
This work was supported in part by the Office of High Energy Physics of the U.S.\ Department of Energy under contract DE-AC02-05CH11231 and by the Natural Sciences and Engineering Research Council of Canada.

\newpage

\bibliography{butauNotes}

\end{document}